\documentclass[a4paper]{article}

\usepackage[english]{babel} 
\usepackage[utf8x]{inputenc}
\usepackage[T1]{fontenc}
\usepackage{amsmath}
\usepackage{multirow}
\usepackage{rotating}
\usepackage{graphicx}
\usepackage{url}
\usepackage{listings}
\usepackage{color} 
\usepackage{appendix}

\definecolor{backcolor}{rgb}{0.9,0.9,0.95}
\lstdefinestyle{code}{
	backgroundcolor = \color{backcolor},
    basicstyle=\footnotesize
}
\usepackage[a4paper,top=3cm,bottom=2cm,left=3cm,right=3cm,marginparwidth=1.75cm]{geometry}

\usepackage{amsmath}
\usepackage{graphicx}
\usepackage[colorinlistoftodos]{todonotes}
\usepackage[colorlinks=true, allcolors=blue]{hyperref}

\usepackage{tikz}
\usetikzlibrary{arrows}

\title{Quantitative analysis of approaches to group marking}
\author{Hugh Harvey\footnote{These authors contributed equally} \footnote{hh15160@bristol.ac.uk} , James Keen\footnotemark[1] \footnote{24jkeen@gmail.com} , Chester Robinson\footnotemark[1] \footnote{cr15169@bristol.ac.uk} , James Roff\footnotemark[1]  \footnote{james.roff.95@gmail.com, Corresponding Author} ~and Thilo Gross\footnote{thilo2gross@gmail.com}\\
{\small University of Bristol, Department of Engineering Mathematics, Woodland Road, Bristol, UK}}

\begin{document}
\maketitle

\begin{abstract}
Group work, where students work on projects to overcome challenges together, has numerous advantages, including learning of important transferable skills, better learning experience and increased motivation. However, in many academic systems the advantages of group projects clash with the need to assign individualised marks to students. A number of different schemes have been proposed to individualise group project marks, these include marking of individual reflexive accounts of the group work and peer assessment. Here we explore a number of these schemes in computational experiments with an artificial student population. Our analysis highlights the advantages and disadvantages of each scheme and particularly reveals the power of a new scheme proposed here that we call pseudoinverse marking. 
\end{abstract}

\section{Introduction}
Group projects, where a small group of students tackles a challenge, can enhance the learning experience~\cite{Abelson1986,HmeloSilver2004}. A central idea of this form of teaching is that the students help each other and learn to overcome problems together. This is beneficial as it allows students to develop transferable skills in project management and leadership, besides technical skills~\cite{HmeloSilver2004,Davies2009}. 

To an assessor, group projects can present a considerable challenge~\cite{Fellenz2006,Davies2009}. Many academic systems require the assessor to assign an individualised mark to each student participating in the course. These marks should not only be fair and unbiased, but should also take the transferable skills that the student acquired into account~\cite{Jackel2017}. While the technical progress made by a group is generally easy to assess, e.g.~on the basis of a project report, it is more difficult to judge the technical and collaborative skills of individual members of a group. In a typical setting, each student works on several shorter projects with different groups. The challenge is then to compute a student's individualised mark from the set of group marks received.

A simple approach is to assign to each student the mean of the marks of the groups in which the student participated. However, as we show below, this has the undesirable effect of reducing the variance of the marks, producing a cluster around the mean. A number of alternatives exist: The assessors can individualise marks based on their own experience of the project or they can ask the students to write reflexive accounts of the project, which are taken into account in the marking~\cite{Dyment2011}. 

A recent review of group marking~\cite{Dijkstra2016} noted that the most common approach to individualised marking is peer assessment, where students are asked to mark the other members of the groups they participated in, by assigning actual marks, e.g.~\cite{Abelson1986,Fellenz2006}, or grouping them in categories, e.g.~\cite{Lejk2001}. It has been noted that peer marking is an effective tool if used appropriately, but there can be some flaws when dealing with students who are either particularly generous or harsh with their assessment of peers~\cite{goldfinch1994}.

Other options include self-assessment, where you ask the student within the group to perform an additional piece of work and assign a portion of their mark based on that. It is noted that, while self-assessment is a simple approach, it often deviates from the markers assessment significantly, partly due to factors beyond the scope of the project~\cite{sharp2006deriving,sunol2016peer}.

An interesting scheme proposed in~\cite{ko2014peer} reveals another alternative, here called Iterated Individual Weighting Factor (IWF-it). This method involves peer assessment as above, but then adjusting the weighting of each group member's opinion depending on how close to the true group mark they predicted for their group. This approach does attempt to discount potential bias in the peer assessments by inaccurate individuals, but is computationally expensive. 

There has been an attempt to simplify the process of marking group projects by the Australian Learning and Teaching Council. This resulted in a software package called \textnormal{$SPARK^{PLUS}$} (Self and Peer Assessment Resource Kit)~\cite{freeman2002spark}. This software has been tested in a university-level engineering setting~\cite{wu2014implementation}. The test showed that, while the idea was sound, a high proportion of the test subjects were dissatisfied with the programs attempt to distribute the group marks.

Here, we consider the effectiveness of different approaches through quantitative analysis of fairness. Our analysis highlights two promising approaches: First, a particular peer assessment method in which each group member is asked to assess the value of each member of the group, in order to then fairly divide the marks between them. While there is a risk with this approach that marks can be manipulated by coordinated responses, it otherwise yields fair results from a transparent and easily understood procedure. Second, pseudoinverse marking can be implemented to eliminate the need for peer assessment. This approach uses the marks of each individual student for a series of projects and then computes the best estimate of the mark that the student should receive. This approach avoids the risk of strategic coalitions and the administrative effort of peer assessment at the cost of transparency. 

\section{Methods}
In the following, we test various marking schemes by applying them to a virtual student population. 
We numerically generate populations of $N$ students. Each student $i$ is assigned a number which represents their ``ideal mark'' $q_i$, i.e.~the mark that they would receive under an optimal marking scheme. We draw the ideal marks for the virtual student population from a Gaussian distribution with mean mark of 60 and standard deviation of 12, which is consistent with assumptions of the UK academic system. 

After the virtual population has been created, we assume that each student undertakes $p$ projects in groups of size $m$ over the course of the unit. The unit mark for each student is then found as some function of the marks of the groups the student participated in, and potentially some other information, such as peer assessment. 

Without much analysis, it is intuitive that having larger groups (greater $m$) makes it harder to determine accurate individualised marks, whereas having more group marks for each student (greater $p$) makes it easier. We verified that this is indeed the case for all schemes considered. For all but the last scheme studied in this paper the effect of participation in more groups is very predictable as it just leads to an averaging. For the analysis presented here, we vary group size and project per student together and consider specifically the case $m=p$, which allows us to present results more concisely.  

In our our virtual population simulation, we assume that the group mark $w_j$ for group $j$ is the mean of the ideal marks of the group's participants, which we can mathematically express as 
\begin{equation}
w_j = \frac{\sum_i M_{ij} q_i}{n_j},
\end{equation}
where $n_j=\sum_i M_{ij}$ is the number of participants of project $j$, and   $\bf M$ is the participation matrix which is defined by 
\begin{equation}
M_{ji} = \left\{ \begin{array}{l l} 1 & \mbox{if student $i$ participated in project $j$,} \\
   0  & \mbox{if student $i$ did not participate in project $j$.} \end{array} \right.
\end{equation}

For illustration consider the matrix 
\begin{equation}
\mathbf{M} = \left (\begin{matrix} 1&1&0&0\\ 0&0&1&1\\ 1&0&1&0\\ 0&1&0&1 \end{matrix}\right).
\end{equation}
This matrix describes the partitioning of 4 students into 4 groups, such that student 1 and student 2 participate in group 1 (first row), student 3 and student 4 participate in  group 2 (second row),
student 1 and student 3 participate in group 3 (third row) and student 2 and student 4 participate in group 4 (fourth row).

Once the group marks have been determined we apply a set of different marking schemes (explained below) to assign individualised marks $x_i$ to the students. We analyse the accuracy of the marking schemes first by considering their performance in a scenario where $n=52$ and $m=4$. For this scenario, we can draw scatter plots showing the individualised mark $x_i$ of a student as a function of their ideal mark $q_i$. We furthermore study the accuracy as a function of the variables $n$ and $m$. For this comparison, we quantify the accuracy in terms of the maximal absolute error 
\begin{equation}
E_{\rm max} = {\rm max}_i |q_i-x_i|,
\end{equation}
the mean absolute error
\begin{equation}
E_{\rm mean} = \sum_i \frac{|q_i-x_i|}{n},
\end{equation}
and the root mean square error
\begin{equation}
E_{\rm rms} = \sqrt{\sum_i \frac{(q_i-x_i)^2}{n^2}}.
\end{equation}
Among these, the mean absolute error provided a measure of the overall accuracy, whereas the maximum error provides an estimate of ``unfairness'' by focussing on the most extreme deviation across the group. The mean square error is an intermediate between these two extremes, averaging over the population but assigning a higher weight to individual marks that deviate strongly.  

\section{Results}
We now compare six different schemes for generating individualising marks. Starting from the simplest and building up to the most mathematically complex. The first is simply to assign the group mark to all the group members (Sec.~\ref{secSOPP}), which provides a baseline or null-model against which other methods can be judged. We then consider using additional information such as reflexive accounts (Sec.~\ref{secRA}), before proposing an improvement on this scheme that we call mark-adjusted reflexive 
accounts (Sec.~\ref{secMRA}). Subsequently, we consider two methods for peer marking, normalised peer assessment (Sec.~\ref{secNPA}) and peer ranking (Sec.~\ref{secPR}), a scheme we propose based on an approach that has been proposed for sports rankings~\cite{Park,Motegi}. Finally, we propose a mathematical method that aggregates results from different projects to infer individualised marks using the Moore-Penrose pseudoinverse (Sec.~\ref{secPI}).   

\subsection{Self-organised peer pressure (SOPP) \label{secSOPP}}
In our simplest scheme, the final mark for a student is the average of the marks of the groups that the students participated in 
\begin{equation}
x_i = \sum_j \frac{M_{ij} w_j}{n_i}, 
\end{equation}
where $n_i = \sum_j M_{ij}$. We call this scheme the self-organised peer pressure (SOPP) method because every student has a direct interest in the success of their groups, whereas other marking schemes might create secondary objectives such as to improve standing within the group 
(possibly at a cost to other group members or group success) to optimise outcomes from peer assessment.    

We note that even the very simple SOPP method leads to individualised marks unless multiple students participate in exactly the same groups. 

The disadvantage of this method is that it leads to a ``regression to the mean''-type of effect. Plotting the expected assigned versus the ideal mark for a group of students (Fig.~1) shows that the best students receive on average less than their ideal mark whereas weaker students receive more than their ideal mark. This result is intuitive as, under this scheme, good students suffer from being grouped with students that are on average weaker than themselves, whereas weak students benefit from being groups with students who are on average stronger than themselves. 

Considering the effect of students participating in more, larger projects (Fig.~2) shows that average errors saturate. However, this saturation occurs at a relatively high level where the root mean square error reaches almost 10 percentage points, which corresponds to a whole degree classification in the British academic system.  Perhaps more worrying is that the maximum error is high and keeps increasing, indicating that this marking scheme is increasingly unfair to individual students.

In summary, the advantage of the SOPP scheme is that it is very easy and intuitive. This makes the scheme easy to apply and marks are found by a highly transparent procedure. Furthermore, it creates incentives that are well aligned with the spirit of project work. For each student, the only way to improve their marks is to improve the mark of the groups they participate in. This mimics a typical workplace situation where ultimately the success of a project matters rather than the contributions of individual team members.    

The good alignment of the assigned mark and project goals is undermined by the large error in the marks, which is of the same order of magnitude than the range of the marks returned by the method. One may be tempted to rescale the marks to address the regression-to-the-mean, resulting in a narrow distribution of marks. However, such a rescaling would exacerbate the individual errors to levels that could be judged unacceptable.     

\subsection{Reflexive accounts (RA) \label{secRA}}
A common approach to individualise marks is to pair group projects with individual assignments such as reflexive accounts on the project. Here we assume that for every group project a student participates in, they also carry out an individualised assignment.
We assume that the assessor does not arrive at the student's ideal mark when marking the individualised component. This represents the student performing better or worse for a reflexive report than in their report, and possible subjectivities in the assessors marking. The assessor instead arrives at the students ideal mark with an error, drawn from an uniform distribution with a range of $\pm 16$.

In the simplest case, considered in this section, the mark that the student receives in a given project is then found as a linear combination of the group mark and the mark for the individual assignment. Based on personal experience and some preliminary trials we focus on the case where the group mark enters with weight $\alpha = 0.7$ and the individualised component enters with weight $ 1 - \alpha = 0.3$. This means that, taking multiple projects into account the final mark of a student is computed as 
\begin{equation}
x_i = \frac{1}{n_i}\sum_j \alpha w_j M_{ij} + (1 - \alpha) y_{ij} = \frac{0.7 \cdot \sum_j M_{ij} w_j + 0.3 \cdot \sum_j y_{ij}}{n_i},
\end{equation}

where $y_{ij} \left( q_i \right)$ represents the mark assigned to student $i$ on project $j$ for the reflexive piece of work, as a function of their ideal mark  $q_i$. As the final mark is to a significant proportion based on the ideal mark, we expect better overall performance of this scheme. The regression-to-the-mean that we identified as the major problem of the SOPP method is ameliorated but not eliminated (Fig.~1). Also considering the participation in larger groups shows a quantitatively better, but qualitatively similar picture to the SOPP method. 

In summary, the RA approach offers few surprises. It uses a linear combination of the SOPP method with the ideal result and hence produces results that interpolate between the SOPP outcome and the ideal outcome. The advantage of the RA scheme is that we can somewhat ameliorate the poor performance of the SOPP scheme, while largely maintaining transparency mathematical simplicity. However, the improvement comes at the cost of abandoning the spirit of project work to some extent as the project outcome is supplemented by a secondary objective which is an individual written assignment. We thus lose some of the beneficial alignment of the project with real-life workplace scenarios and introduce significant non-project workload for both the student and the assessor. 

Changing the weighting of group and individualised components in the marking gives some control over the balance between the advantages of project work and marking fairness. 
What makes this scheme somewhat unsatisfactory is that the trade-off remains linear. Introducing a small individualised component only marginally improves mark accuracy, while significantly improving accuracy requires us to sacrifice most of the advantages of project-based assessment.  

\subsection{Mark-adjusted reflexive accounts (MRA) \label{secMRA}}
We now explore an alternative use of marks for reflexive account, where the final mark is not entered linearly but are instead used to judge the individual's contribution to the project. We assume again that the mark received $y_i$ for the reflexive account is based on the student's ideal mark $q_i$. The same uniform distribution with range of $\pm 16$ is used to represent the reflexive marking error. We then estimate the contribution student $i$ to project $j$ by
\begin{equation}
c_{ij}=\frac{y_i}{\sum_k y_k M_{kj}}.
\end{equation}
The total number of mark points $t_j$ that we can distribute for project $j$ is the project mark $w_j$ times the number of students involved in the project $n_j=\sum M_{ij}$, i.e.
\begin{equation}
\label{eqTotalPoints}
t_j = w_j n_j.
\end{equation}
Instead of distributing these points equally we now partition them  according to the estimated contribution to the project, such that student $i$ receives for project $j$ 
\begin{equation}
r_{ij}=c_{ij}t_j = \frac{w_j y_i \sum_k M_{kj}}{\sum_k y_k M_{kj}}.
\end{equation}
The final mark for the student after participating in multiple projects is then
\begin{equation}
x_i = \frac{\sum_j M_{ij} r_{ij}}{\sum_j M_{ij}}.
\end{equation}
\begin{figure}[ht]
  \centering
    \includegraphics[width=0.49\textwidth]{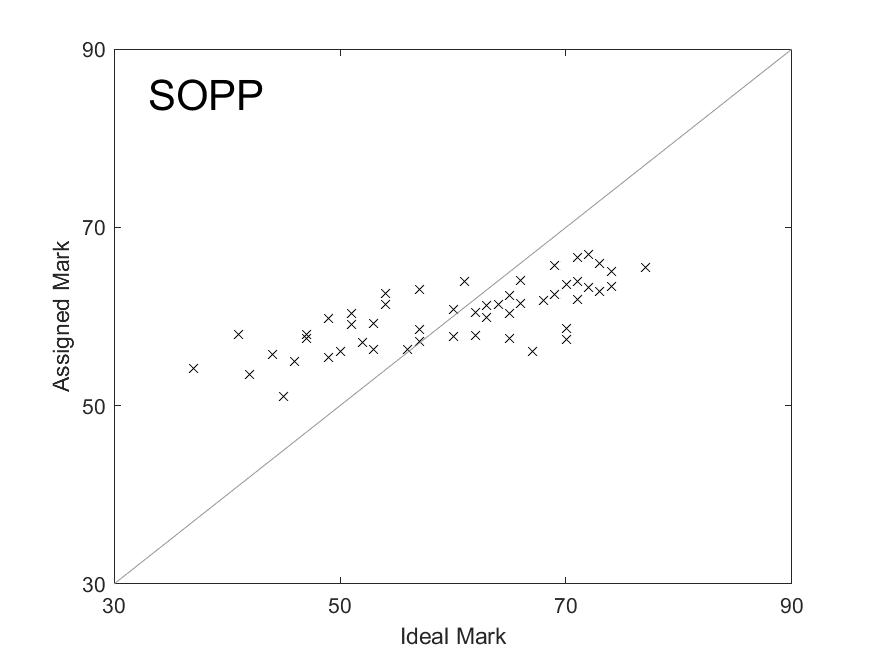} \hfill
    \includegraphics[width=0.49\textwidth]{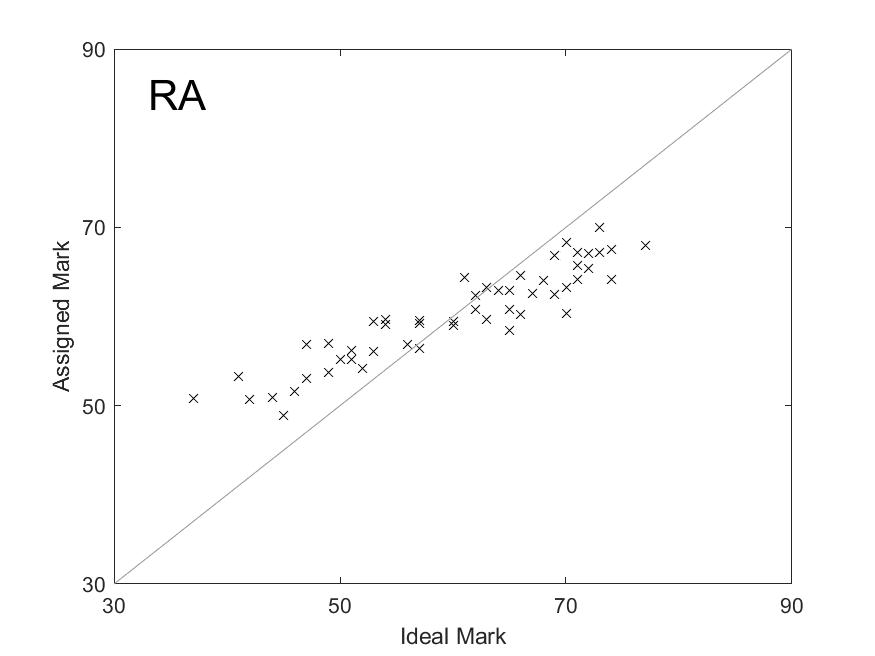}
    \includegraphics[width=0.49\textwidth]{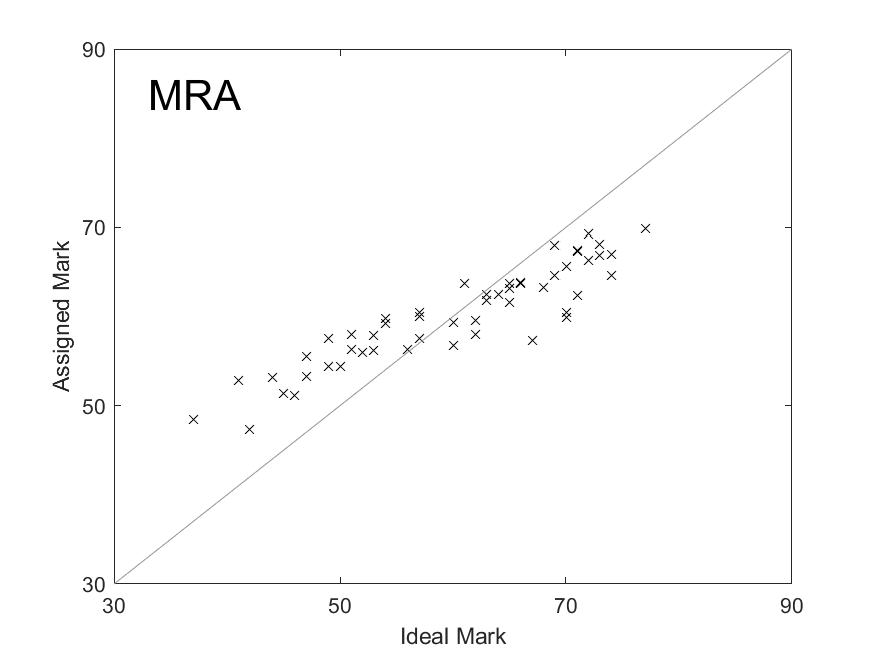} \hfill
    \includegraphics[width=0.49\textwidth]{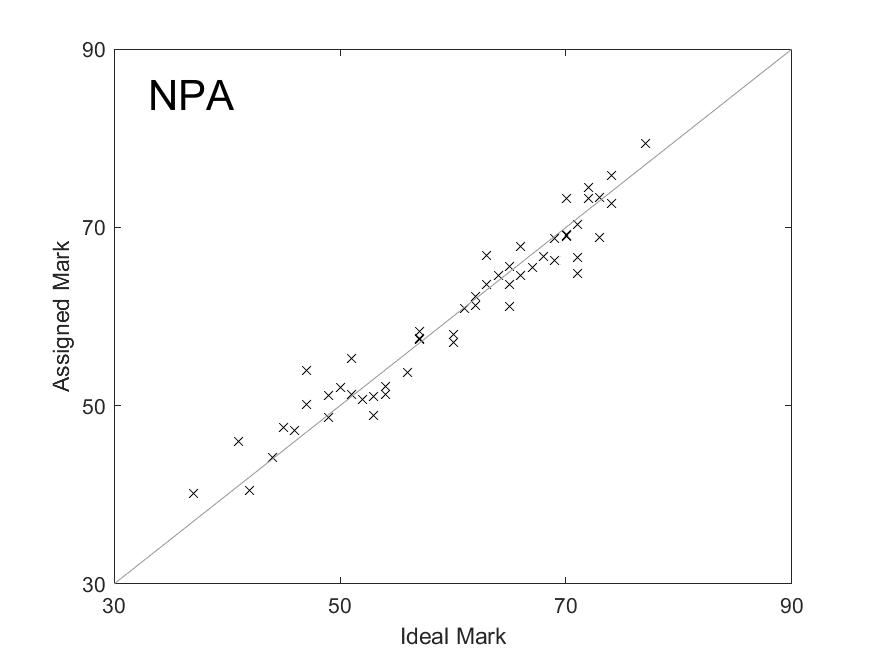}
    \includegraphics[width=0.49\textwidth]{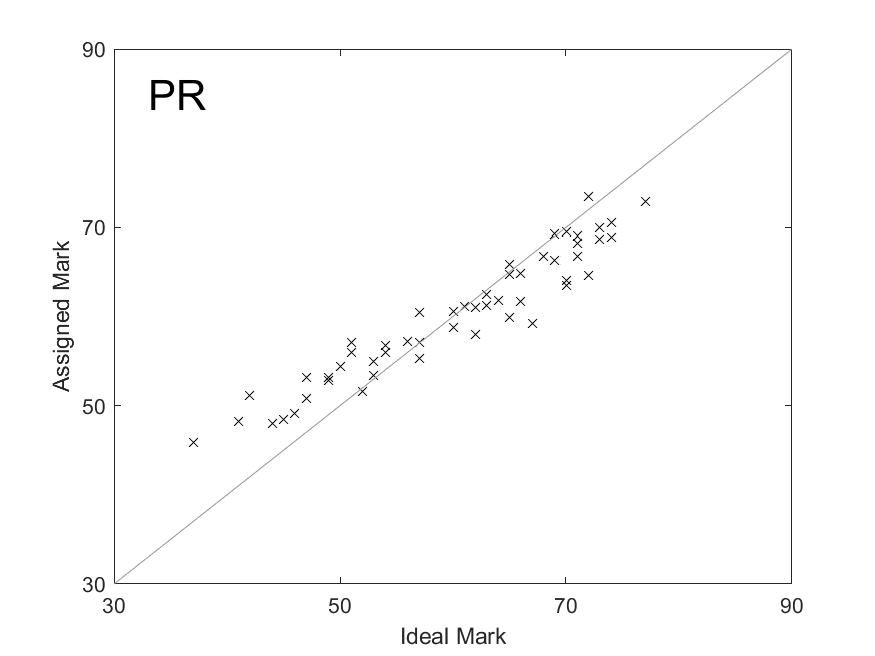} \hfill
    \includegraphics[width=0.49\textwidth]{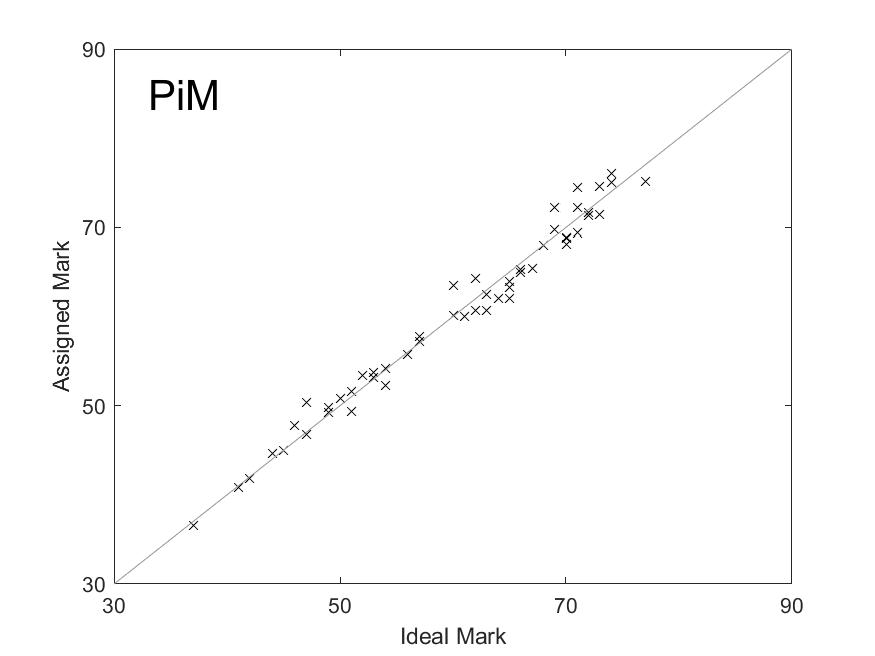}
    \caption{Scatter plot of assigned mark against ideal mark for various marking schemes. Shown are results for Self organised peer assessment (SOPP - top left), Reflexive accounts (RA - top right), Mark-adjusted reflexive accounts (MRA - middle left), Normalised peer assessment (NPA - middle right), Peer ranking (PR - bottom left) and Pseudoinverse marking (PiM - bottom right). Diagonal lines indicate the ideal distribution of assigned marks. Pseudoinverse marking is the best distribution of marks. ($N = 52$, $m = 4$, and $p = 4$)\label{Fig-1}}
\end{figure}
\begin{figure}[ht]
  \centering
    \includegraphics[width=0.49\textwidth]{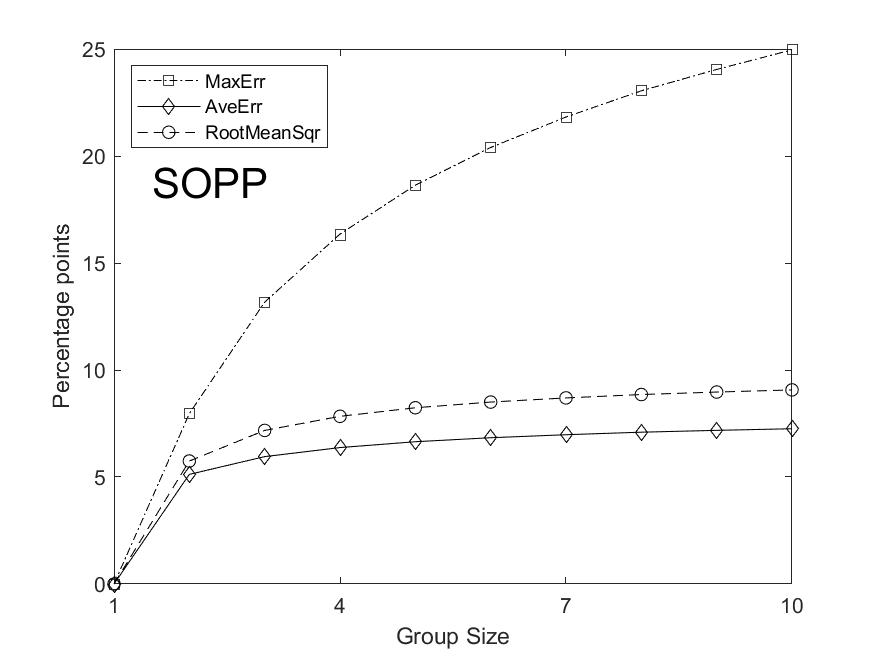} \hfill
    \includegraphics[width=0.49\textwidth]{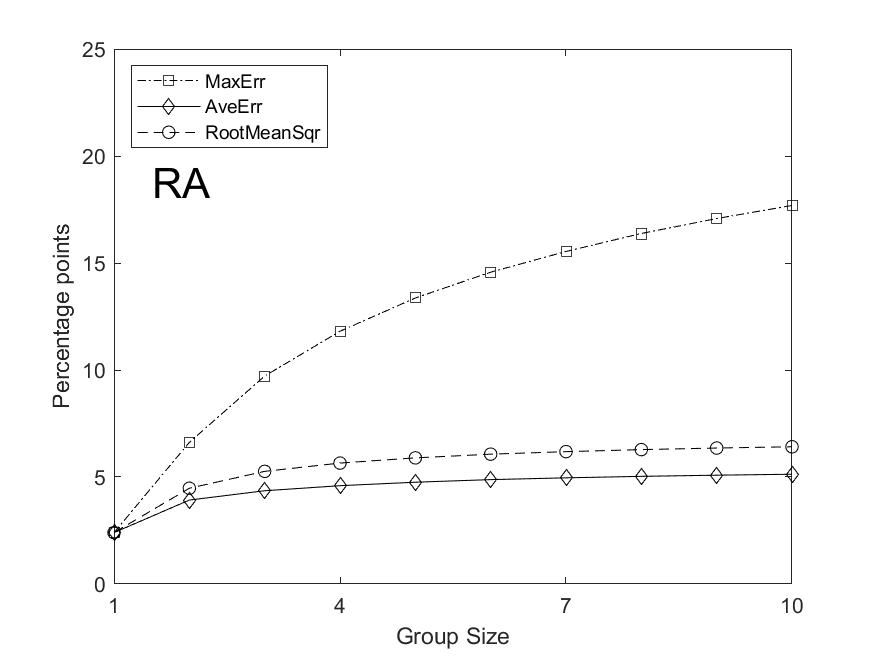}
    \includegraphics[width=0.49\textwidth]{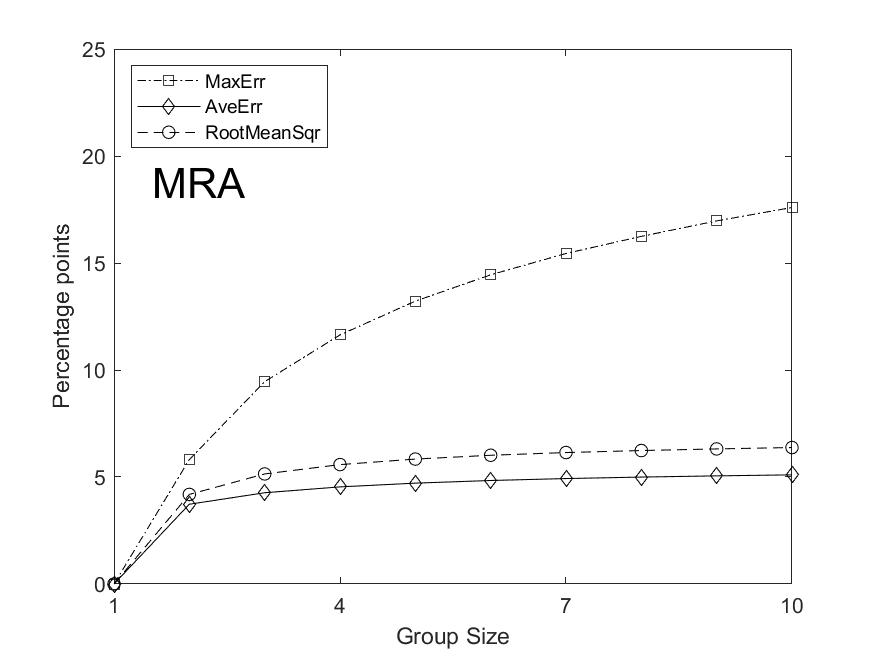} \hfill
    \includegraphics[width=0.49\textwidth]{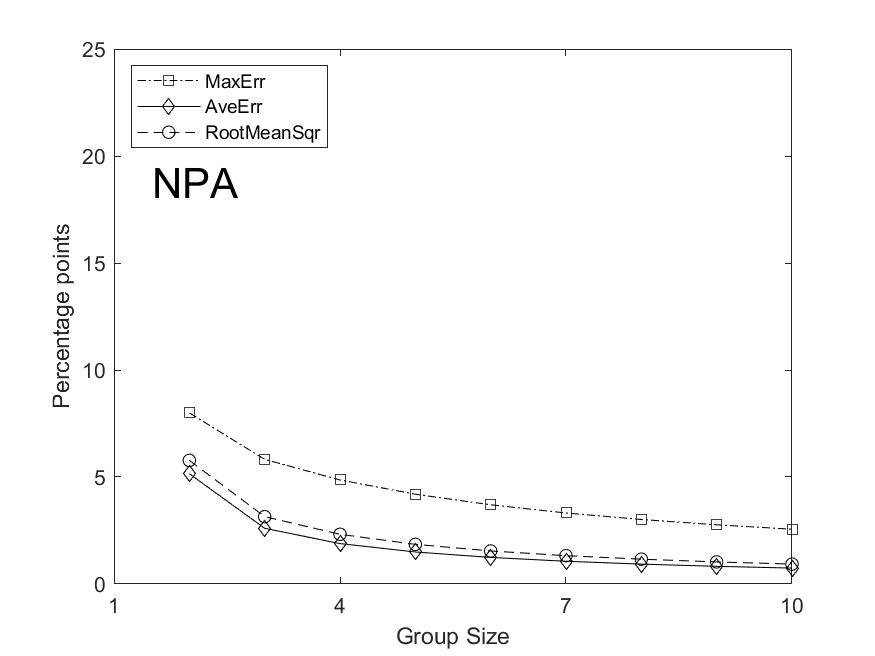}
    \includegraphics[width=0.49\textwidth]{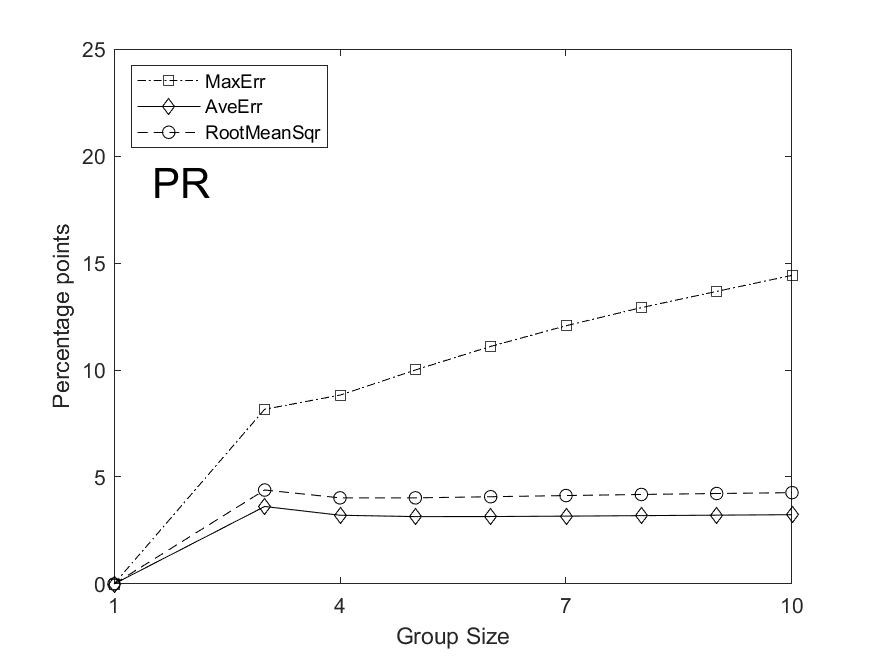} \hfill
    \includegraphics[width=0.49\textwidth]{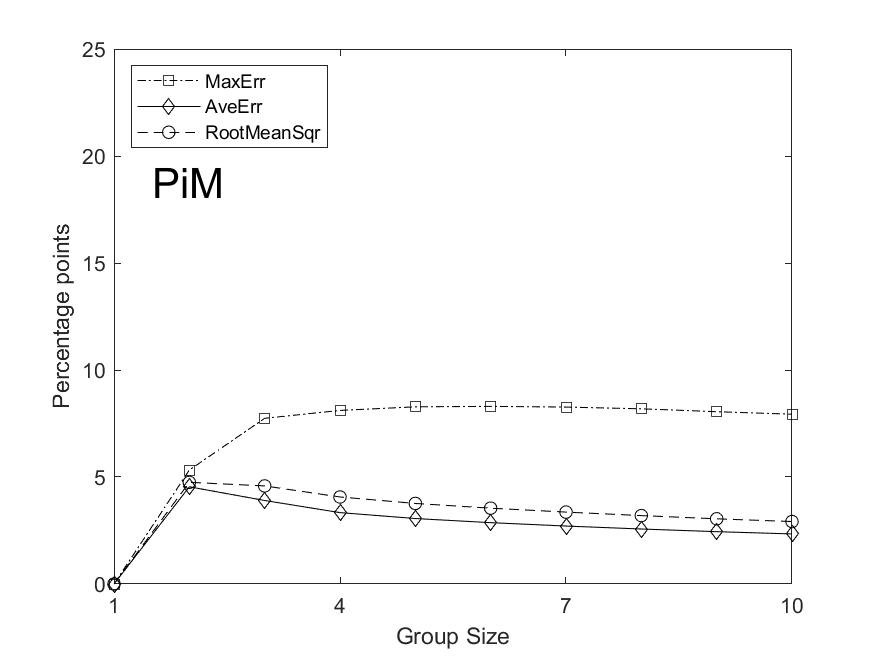}
    \caption{Plot of the different measures of error against different group sizes $m$ for various marking schemes. Shown are results for Self organised peer assessment (SOPP - top left), Reflexive accounts (RA - top right), Mark-adjusted reflexive accounts (MRA - middle left), Normalised peer assessment (NPA - middle right), Peer ranking (PR - bottom left) and Pseudoinverse marking (PiM - bottom right). Normalised peer assessment and Pseudoinverse marking show interesting trends as using larger group sizes reduces all three errors.\label{Fig-2}}
\end{figure}
Considering the outcomes of the computational experiment we see that good students still receive marks that are systematically less than their ideal mark (Fig.~1). Compared to the simpler RA scheme the performance of MRA is almost exactly identical. This is also confirmed when considering different group sizes and project iterations (Fig.~2).
A small difference exists for very small group sizes, but this is mostly cosmetic. Consider that, under the MRA scheme, the reflexive account does not affect the marking at all. Thus any error made in the marking of the reflexive account does not affect the student's mark. So this difference appears due to specific assumptions of the computation experiment and is of little practical relevance.   

Perhaps more significant is another advantage of the MRA scheme that is not considered in the computational experiment. This is the case where one student in an otherwise well-performing group does not engage with the project at all. 

Consider the (somewhat extreme but not unheard of) example where a student does not do any work for the project or the reflexive account, but the project is still marked as 60. Under the assumptions made in this report the average ideal mark for the other 3 students on the project would have to be 80 to achieve this result while compensating for the work not done by the defecting student. Assuming that the 3 engaged students achieve on average the same level of marks in their reflexive accounts their mark under the RA scheme would be 
\begin{equation}
0.7\cdot 60 + 0.3 \cdot 80 = 42 + 24 = 68 .
\end{equation}
While the defecting student receives
\begin{equation}
0.7\cdot 60 + 0.3 \cdot 0 = 42 , 
\end{equation}
which in the British system would mean that the defecting student still passes while the very strong engaged students only receive an upper second class mark. By contrast, in the MRA scheme the engaged students would receive on average their ideal mark of 80, a solid first class result, while the defecting student receives a zero mark. 

Even if the defecting student decides to invest effort in the reflexive account, but not the project, the student would need to achieve a mark of 51 in the reflexive account to receive the same mark of 42 that would be assigned under the RA scheme. We note that the MRA scheme can mean that students can fail although both their reflexive account and their overall group mark are sufficient to pass. However, this is unlikely to happen unless a student only participates in groups where all others receive vastly higher marks on the reflexive accounts.   

In summary, the MRA scheme is slightly more complex than the RA scheme, but the calculations are still simple enough to be carried out very quickly with pen and paper, a calculator or a simple spreadsheet. On average, the improvement of performance over the RA scheme does not seem to justify the extra complexity of the MRA. However, MRA is superior in the extreme case where students do not engage with the project at all. Because the scheme is still quite transparent, this means that MRA can create strong incentives for the students to engage with the projects. We caution however that it is prudent to be conservative in the marking of reflexive accounts if this scheme is used and avoid extreme marks unless they are clearly indicated.  

\subsection{Normalised peer assessment (NPA) \label{secNPA}} 
In peer assessment the students assess the performance of each member of their group. 
The immediate advantage of this method is that information on the relative contributions to the project is sourced directly from the students involved. The obvious drawback is that the students are given a way to influence their mark other than delivering a good project outcome. There is thus a risk that students use peer assessment strategically to maximise their own marks rather than to provide genuine information about project contributions. 

A variety of different peer assessment schemes have been proposed to maximise the advantages and minimise the disadvantages of this approach~\cite{PeerAssessment}. To gain insights into the student's ability to manipulate their marks, one can distinguish between strategies that students can implement for themselves and those that require the formation of coalitions with other students. An example of the former case would be a student could give themselves a high mark and/or all other group members a low mark to maximise their outcome. An example of the latter case would 3 members of a group of four conspiring to increase their marks at the expense of the fourth student. 

The topic of coalition formation in groups is very complex and subject of active research. 
However, a simple argument can be made to show that peer assessment can fail if multiple students conspire to mark others strategically. Consider the (not very far fetched) situation where three members of a four-person group conspire against the fourth student. In this case the three conspirators could mark each other highly while assigning a zero mark to the fourth student. In the absence of other sources of information, this situation is indistinguishable from a scenario in which one of the students did not engage with the project at all. Thus no marking scheme can deliver a satisfactory outcome in both of these scenarios at the same time. 

Here we particularly consider a normalised peer marking scheme~\cite{PeerAssessment}. In this scheme every member of a group marks all other members, but not themselves. Not allowing students to mark themselves prevents students from inflating their own mark, the most direct form of mark manipulation. 

In the next step the marks assigned by each student are normalised. For example, let $S_{ki}$ be the mark that student $k$ assigned to student $i$. We can then compute the normalised mark
\begin{equation}
s_{ki}= \frac{S_{ki}}{\sum_i S_{ki}},
\end{equation}
the normalised marks $s_{ki}$ are in the range [0,1] and reflect the student $k$'s opinion of the proportional contribution of the other group members.
Taking only these normalised marks into account prevents students from inflating their own contribution by marking all others contributions lowly. 

To determine the mark $r_{ij}$ that student $i$ receives for the project $j$, we first compute the total mark points as in the MRA scheme above (Eq.~\ref{eqTotalPoints}) and then distribute these points proportionally to the normalised marks a student has received 
\begin{equation}
r_{ij} =\frac{t_j \sum_k s_{kj}}{\sum_i M_{ij}}, 
\end{equation}
where $t_j$ is the total number of mark points that can be distributed according to (Eq.~\ref{eqTotalPoints}) and the denominator is the number of students in the group.

In the numerical simulation of this method, each student is assumed to rank another student based on their ideal mark with some random error. This random error represents students not accurately judging the ability of other group members. Like for similar errors for other marking schemes, an uniform distribution with a range of $\pm 16$ was used. Under these assumptions the method achieves good results, particularly there is no systematic bias against stronger students (Fig.~2).  Unlike all other schemes in this report, the error of this method decreases when group size is increased. This is intuitive as the assigned mark builds is computed from a larger number of observations, and thus profits from a ``Wisdom of the Crowd'' effect~\cite{Wisdom}. 

While the performance of this method in our test is very good, the clear drawback of the method is that it is vulnerable to manipulation by coalition formation. One could argue that this can be somewhat mitigated by adjusting the mark if indications of a coalition formation within a group exists (e.g.~from observation of the group work). However, allowing for such adjustments to some extend defeats the purpose of having a transparent procedure. We therefore recommend this method particularly for projects with large group sizes, where the method performs particularly well, and the impact of small coalitions within the group is lesser. 

\subsection{Peer ranking (PR) \label{secPR}}
We now turn to two unusual methods. For sports rankings, network-based approaches have recently received attention~\cite{Park,Motegi}. A similar approach can be taken to peer marking. Instead of marking each group member with a precise mark students rank the other members of their group. The ranking consists of a pairwise comparison of the perceived contribution of every pair of other members. 
We can interpret each such ranking as a link in the network leading from the weaker to the stronger student. (see Fig.~\ref{Rank_Graph_3} for an example) 
\begin{figure}[ht]
  \centering
  \begin{minipage}[ht]{0.49\textwidth}
  \centering
    \begin{tikzpicture}

	\tikzset{vertex/.style = {shape=circle,draw,minimum size=1cm, thick}}
	\tikzset{edge/.style = {->,> = latex , thick }}

	\node[vertex] (a) at  (0,0) {\textbf{A}};
	\node[vertex] (c) at  (0,-2.5) {\textbf{C}};
	\node[vertex] (b) at  (5,0) {\textbf{B}};
	\node[vertex] (d) at  (5,-2.5) {\textbf{D}};


	\draw[edge] (a)  to[bend left] (b);
	\draw[edge] (a)  to[] (c);
	\draw[edge] (a)  to[bend left = 10] (d);
	\draw[edge] (b)  to[] (a);
    \draw[edge] (b)  to[bend right=10] (c);
	\draw[edge] (b)  to[bend left] (d);
	\draw[edge] (c)  to[bend left] (a);
	\draw[edge] (c)  to[bend right=10] (b);
	\draw[edge] (c)  to[] (d);
	\draw[edge] (d)  to[bend left=10] (a);
	\draw[edge] (d)  to[] (b);
	\draw[edge] (d)  to[bend left] (c);

  \end{tikzpicture}
  \end{minipage}
  \hfill
  \begin{minipage}[ht]{0.49\textwidth}
    \begin{displaymath}
		{\bf A} = \left (\begin{matrix} 0&2a&a+b&2b\\ 2b&0&2b&2b\\ a+b&2a&0&2b\\ 2a&2a&2a&0 \end{matrix}\right)
	\end{displaymath}
    \begin{center}\vspace{1.5em}
    Student A: $ B > C > D$ \\
    Student B: $ C > A > D$ \\
    Student C: $ B > A > D$ \\
    Student D: $ B > A > C$
    \end{center}
  \end{minipage}
  \caption{Example of peer ranking. In a set of four students each student forms an opinion about the relative contributions of the other three students (Bottom right). This information can be represented as a directed graph (left), which can be in turn represented as a matrix (top right). The leading eigenvector of this matrix provides an aggreagted measure for the relative contributions.  \label{Rank_Graph_3}}
\end{figure}

Network structures can be encoded in matrices. Here we use a modified adjacency matrix $\bf A$, defined by 
\begin{equation}
A_{ik} = n_a a + n_b b 
\end{equation}
where $n_a$ is the number of times that student $i$ was ranked lower than student $k$, is the number of times that student $i$ was ranked higher than student $k$ and $a$, $b<a$ are constant parameters. An example can be seen in Figure~\ref{Rank_Graph_3}.

The advantage of the matrix notation is that we can now generate marks using a spectral approach, similar to the famous PageRank algorithm~\cite{PageRank}. 
For this purpose, we compute the leading eigenvector $\boldsymbol{v}$ of $\bf A$
and normalise it such that $\sum v_n = n_j$. 

After the normalisation, the entries of the eigenvector are in the range $[0,n_j]$
with a mean $1$. The $n$-th entry $v_n$ is a proxy for the relative contribution for student $n$ in the group.

We then use the eigenvector entries to personalise the marks such that the final mark that student $i$ receives for the project $j$ is 
\begin{equation}
r_{ij} = w_j (\alpha + (1-\alpha)v_n)  
\end{equation}
where $n=n(i)$ is index assigned to student $i$ within group $j$ and $\alpha$ is another scalar parameter used to control the weight attributed to the peer ranking. 

Based on preliminary tests we used the parameters $a=0.25$, $b=1$ and $\alpha=0.65$. We assumed that students rank each other student in order of a perceived mark, which is the ideal mark and a random error. Like for the normalised peer assessment method (Sec.~\ref{secNPA}), this random error is drawn from an uniform distribution with range $\pm 16$. These perceived marks are used to rank each other student, and then only the set of ranked lists from each student is used. 

The results of our computational experiment show that these choices of parameters receive a relatively low absolute error, but has a slight systematic bias favouring weaker students. This bias could be reduced by reducing $\alpha$, albeit at the cost of increasing non-systematic error. Individual error appears primarily in groups of equal or almost equal strength where the ranking method amplifies the small differences. One may suspect that this is a lesser problem in reality where students are not able to rank each other perfectly based on tiny differences. However, a detailed investigation of this point would likely require a study of ranking behaviour of real students, which is beyond the scope of the present paper. 

In summary, the peer ranking scheme, performs slightly worse than the NPA method (particularly in case of large group sizes). It is also considerably less transparent. We therefore judge that the NPA method will be superior in most cases. 

\subsection{Pseudoinverse Marking (PiM)\label{secPI}}
All methods considered so far were applied to one project group at a time. However, in a setting where students participate in multiple projects additional information can be gained by taking into account how they perform in groups of different composition. 

We continue working on the assumption that the mark of a project report reflects the average ideal mark of the participants in the project, i.e.
\begin{equation}
w_j = \frac{\sum_i M_{ij} q_i}{n_j}.
\end{equation}
In vector notation this can be written as 
\begin{equation}
\boldsymbol{w} = {\bf Q} \boldsymbol{q},
\end{equation}
where the matrix $\bf Q$ is defined by $Q_{ij}=M_{ij}/n_j$. In marking, we have determined the project marks $\boldsymbol{w}$, and we know the $\bf Q$ as it follows from the partitioning of students into groups. Our aim is to compute the ideal marks $\boldsymbol{q}$ that the students should receive. 

Formally we can compute $\boldsymbol{q}$ by multiplying $\bf Q^{-1}$ the inverse of $Q$ which yields 
\begin{equation}
\label{deconv}
\boldsymbol{q} = {\bf Q^{-1}} \boldsymbol{w}.
\end{equation}
If students were partitioned into groups such that $\bf Q$ is invertible, this relationship should yield the desired marks exactly. 

In typical cases another slight complication arises because common ways of dividing students into groups lead to singular matrices $\bf Q$ such that an inverse does not formally exist. In this case, the method can still be applied if we replace the inverse with the Moore-Penrose pseudoinverse~\cite{pseudo}. 
If each student participates in only a single project, then calculating the pseudoinverse of $\mathbf{Q}$ corresponds to implementing the SOPP method (Sec.~\ref{secSOPP}). Clearly, this is undesirable so the number of projects each student participates in must be at least the number of students in the group.
In an ideal case, every student completes multiple projects and interacts with a completely different set of people every time. In this case, Eq.~(\ref{deconv}) can be used to deconvolute the contributions and recover the ideal mark using the pseudoinverse.  

In our numerical experiment, we consider such an case where every student contributes to 4 projects and interacts with $4\cdot 3 = 12$ distinct other students in the process. The assignment resulted in a singular $\bf Q$.
The results (Fig.~1) show that the estimated ideal mark computed with the Moore-Penrose pseudoinverse are in excellent agreement with the ideal marks. The method producing the smallest errors of all methods considered here.

One could criticise that the accuracy of the method relies on our assumption that the project marks achieved are the mean of the ideal marks of the project participants. However, this is actually not so much an assumption as a definition of what is meant by the ideal mark. 

Apart from the high accuracy this method has many other advantages. For example, it does not require any additional information (peer marking, ranking, accounts) from the student and thus eliminates the workload that would otherwise be required to source such information. Moreover, the method is very robust and could be easily implemented, say in a spreadsheet.

\begin{figure}[ht]
  \centering
    \includegraphics[width=0.49\textwidth]{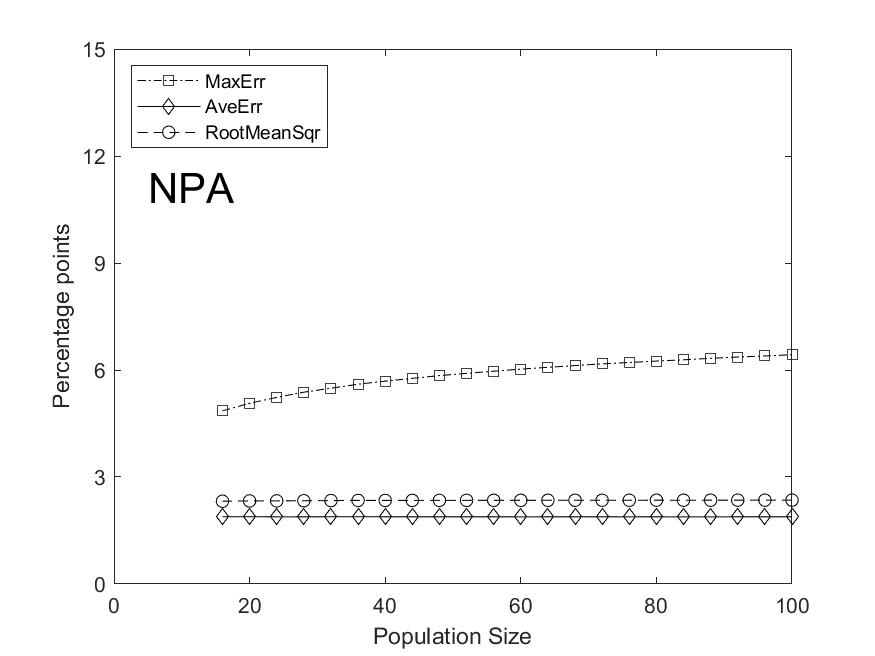} \hfill
    \includegraphics[width=0.49\textwidth]{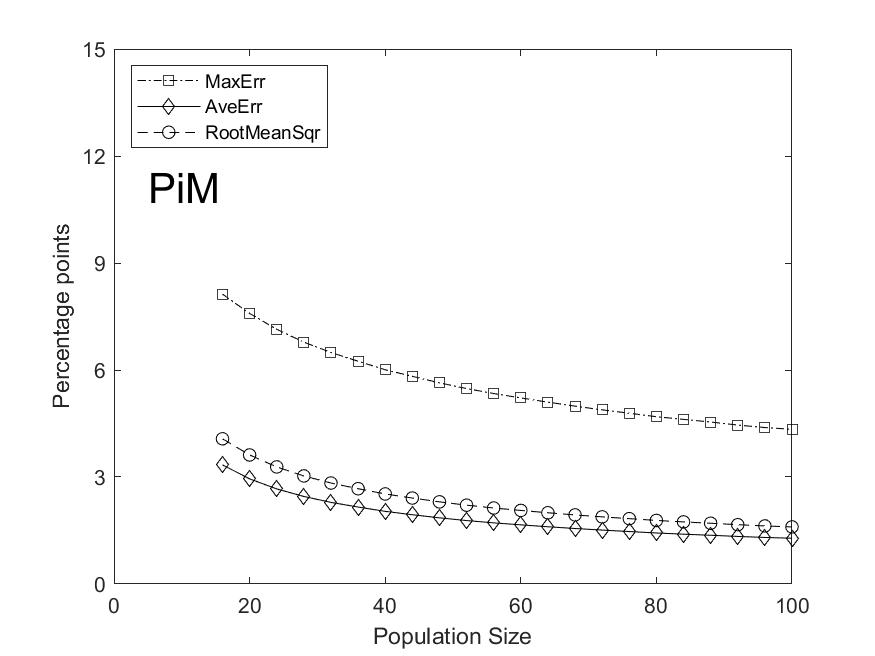}
    \caption{Plot of the different measures of error against different population sizes $N$. Shown are results for Normalised peer assessment (NPA - left)  and Pseudoinverse marking (PiM - right). Results for other methods are similar to NPA and hence have been omitted. Unlike the other methods, Pseudoinverse marking decreases all three measures of error as the population size increases. ($m = 4$, and $p = 4$)\label{Fig-4}}
\end{figure}

Another advantage is that, unlike for any other method, the measured errors all decrease as the population size increases (Fig~\ref{Fig-4}). Intuitively, as the algorithm can draw on more information, the accuracy of the ideal mark predictions should increase. Additionally, for larger populations, more possible student combinations for groups are possible, allowing a more optimal assignment for groups.

A drawback is that the final marks are only available after all projects have been completed, and students may find the method intransparent. Perhaps the main disadvantage is that there is a lower threshold for the number of projects in which a student needs to participate. This de-facto limits the applicability of the method to projects that are carried out in small groups.  

\section{Conclusion}
In this paper we explore several schemes for individualising group project marks in computational experiments using a virtual student population. 
The most suitable scheme depends on the specific circumstances, including the number of students and the number of project each student participates in. 
An overview of the errors for typical settings is shown in Table~1.

\begin{table}[ht]
\centering
\caption{Overview of the errors of various methods for a typical scenario. In a group of 52 students each student undertakes 4 project in groups of 4 people.}
\begin{tabular}{ l | c | c}
\textbf{Method Name} & \textbf{Average Absolute Error} & \textbf{Maximum Error} \\
\hline
Self Organised Peer Pressure &  5.5 & 13.8 \\
Reflexive Accounts & 3.7 & 10.6\\
Mark-Adjusted Reflexive Accounts & 3.5 & 11.3 \\
Scaled Peer Assessment & 1.7 & 4.5 \\
Ranked Peer Assessment & 3.3 & 12.3 \\
Inverse Problem Approach & 1.3 & 1.9\\ 
\end{tabular}
\end{table}

The simplest scheme (SOPP) where students receive the average of the group marks for the projects that they participated in systematically favours weaker students, but has the advantage of not generating additional workload and being very close to a real life workplace situation. Marking of reflexive accounts (RA) did not fully remove the systematic bias from the marks and has the disadvantage of generating significant additional workload. Among the two RA marking schemes, the mark-adjusted reflexive accounts (MRA), proposed here, provides stronger incentive to engage with the project. Normalised Peer Assessment (NPA) resulted in very accurate marks, particularly for projects with many members, but entails the risk of mark-manipulation by coalition formation. A peer ranking scheme (PR) proposed here performed similarly but is significantly less transparent. Finally, pseudo-inverse marking (PiM), also proposed here, achieves very accurate results without the risk of mark-manipulation or additional workload, but requires that the number of projects to which a student contributes is at least as large as the group size in the projects.   

In absence of factors favouring a certain approach our analysis highlights normalised peer assessment as the best scheme for projects with large group sizes and pseudoinverse marking as the best scheme for marking a series of projects carried out in small groups. 

This paper also illustrated how agent based computational models can be used to explore the fairness and accuracy of marking schemes. Here we have used only a very simple model, and plenty of opportunities for improvements and refinements still exist. For example, one could allow the students to allocate their time investment into the project strategically, or build in social dynamics, but these extensions are beyond the scope of the current paper. We hope that in the future more of these analysis will be carried out to yield deeper insights into the mathematical properties of group marking schemes.

\newpage
\bibliographystyle{plain}
\bibliography{ref.bib}

\end{document}